\documentclass[AMA,STIX1COL]{WileyNJD-v2}
\usepackage{moreverb}

\newcommand\BibTeX{{\rmfamily B\kern-.05em \textsc{i\kern-.025em b}\kern-.08em
T\kern-.1667em\lower.7ex\hbox{E}\kern-.125emX}}

\articletype{Article Type}%

\received{<day> <Month>, <year>}
\revised{<day> <Month>, <year>}
\accepted{<day> <Month>, <year>}

\begin{document}
\title{\LARGE Chance-Constrained Model Predictive Control\\ \large A reformulated approach suitable for sewer networks}

\author[1]{Jan Lorenz Svensen*}

\author[2]{Hans Henrik Niemann}

\author[3]{Anne Katrine Vinther Falk}

\author[1]{Niels Kj\o lstad Poulsen}

\authormark{SVENSEN \textsc{et al}}

\address[1]{\orgdiv{Department of Applied Mathematics and Computer Science}, \orgname{Technical University of Denmark}, \orgaddress{\state{2800 Kongens Lyngby}, \country{Denmark}}}

\address[2]{\orgdiv{Department of Electrical Engineering}, \orgname{Technical University of Denmark}, \orgaddress{\state{2800 Kongens Lyngby}, \country{Denmark}}}

\address[3]{\orgdiv{Emerging Technologies}, \orgname{DHI A/S}, \orgaddress{\state{Agern All\'{e} 5, 2970 H\o rsholm}, \country{Denmark}}}

\corres{*Jan Lorenz Svensen, Department of Applied Mathematics and Computer Science, Technical University of Denmark, Asmussens All\'e 303B, 2800 Kgs. Lyngby, Denmark. \email{jlsv@dtu.dk}}

\presentaddress{Asmussens All\'e 303B, 2800 Kgs. Lyngby, Denmark.}

\abstract[Abstract]{In this work, a revised formulation of Chance-Constrained (CC) Model Predictive Control (MPC) is presented.
The focus of this work is on the mathematical formulation of the revised CC-MPC, and the reason behind the need for its revision. 
The revised formulation is given in the context of sewer systems, and their weir overflow structures.
A linear sewer model of the Astlingen Benchmark sewer model is utilized to illustrate the application of the formulation, both mathematically and performance-wise through simulations. 
Based on the simulations, a comparison of performance is done between the revised CC-MPC and a comparable deterministic MPC, with a focus on overflow avoidance, computation time, and operational behavior.
The simulations show similar performance for overflow avoidance for both types of MPC, while the computation time increases slightly for the CC-MPC, together with operational behaviors getting limited.}


\fundingInfo{supported by Innovation Fond Denmark through the Water Smart City project (project 5157-00009B).}


\maketitle


\section{Introduction}\label{sec1}
With the increase in heavy rains in the recent decade\cite{Greg2015}, the operation of sewer systems has become more important. 
In sewage systems\cite{BD11}, there are several objectives for the ideal system operation; control of the flow to the wastewater treatment plant (WWTP), and the avoidance of weir overflows are among some of them. In the previous decades, Model Predictive Control (MPC) has been applied to sewage systems with fair results\cite{GRJ}-\nocite{OMB,HFC,Nadia18,QBJ}\cite{MPB} aiming for the ideal operation. However, the structure of sewage systems is always changing leading to model uncertainties, and the systems being intrinsically driven by rain and dry weather inflows, creating a dependency on the quality of the predictions of those inflows. With the classical MPC being a deterministic method, the presence of uncertainty has not in general been included in the research on MPC for sewer networks.

While the classical MPC is deterministic, MPC methods for handling uncertainty has been developed in the past decades\cite{Mesbah2016}-\nocite{Grosso2014, ECK1,CGP09}\cite{KC16}, but not applied to sewer systems. Collectively these methods are referred to as Stochastic Model Predictive Control (SMPC) and include a wealth of methods. The methods range from finite scenario-based robust approaches\cite{CGP09} to methods based on the probability of constraints being true\cite{Grosso2014, ECK1,KC16}, among other methods\cite{Mesbah2016}.
In this work, we will focus on the method known as chance-constrained MPC (CC-MPC), which previously has been applied to other systems, such as drinking water systems\cite{Grosso2014} with good results. The CC-MPC method utilizes an optimization based on the expected cost of the system, together with probabilistic formulations of the constraints.
The probabilistic formulations are introduced in order to tighten the constraints so that the performance resulting from the controller are feasible within the real constraints with a given probability. 
 
While CC-MPC and other similar SMPC methods can utilize information of the uncertainty to generate constraint tightenings for more robust performances, it does not come without drawbacks. Given that tighter constraints mean the workspace of the controller gets smaller and in worst-case results loss of feasibility for the CC-MPC.
This can happen if the uncertainty is too large or if the desired probability for the constraints to hold are too high, resulting in overlapping constraints and infeasibility.

Another important aspect of MPC design for sewer networks, besides feasibility, is how the overflows from weirs are integrated into the design formulation. Weirs are physical structures with a binary nature of either overflowing or not, giving two different dynamics of the systems to include in the MPC design. Weir overflows are usually integrated by one of three approaches: 1) they are ignored in the formulation and their occurrence means an infeasible scenario. 2) They are integrated into the constraints but excluded from the dynamics. And 3) they are integrated into the dynamics and propagates through the MPC formulation. In this paper, we will consider the third formulation of the weir overflows, however, the CC-MPC mentioned earlier is not suitable for this formulation, due to the overflows being defined by the original constraints (without tightening). The inclusion of overflow into the dynamics leads to another issue with the formulation of CC-MPC.  With the inclusion, the constraints defining the weir elements become intrinsically feasible through the presence of overflow. This results in the probabilistic formulations of the CC-MPC method becomes insensible with probabilities larger or equal to one.

To deal with the above drawbacks of CC-MPC applied to sewer networks, we will in this paper outline and apply a revised formulation of the CC-MPC applicable for sewer networks with weir structures. The reformulation will aim to introduce sensible probabilistic constraints, suitability for the inclusion of weir structures in dynamics, as well as the preservation of the original feasibility of the system.

For the application of the revised CC-MPC formulation will utilize a model of the Astlingen benchmark Network \cite{MMMU}, displayed in Fig. \ref{fig:system}. Furthermore, will the focus of this paper will be given on the revised CC-MPC's performance in view of the classical deterministic MPC's performance when using the third approach to overflow integration.

In the following sections, we will first present the general MPC program for systems with overflows, and then we will discuss and formulate the revised CC-MPC formulation. The paper will end with an applied example and evaluation of the formulated method.

\subsection{Notation}
In this paper, the following notation is utilized. Bold font is utilized to indicate vectors, while a bullet $\bullet$ represents a subset or set of a function's variables. For a stochastic variable \textit{X}, the expectation and variance are denoted $E\{X\}$ and $\sigma^2_{X}$ respectively, while $Pr\{X\leq x\}$ and $\Phi(x)$ is the probability function and cumulative distribution function (CDF) respectively for a given value x. The notation $X \sim F$ indicates that X is following a given distribution F.
The weighted quadratic norm of x is denoted by $||\textbf{x}||^2_A = \textbf{x}^TA\textbf{x}$, while the minimum and maximum of a given function $f(x)$ are denoted $\underline f$ and $\overline f$ respectively. The notation $\Delta T$ and the subscript $k$ indicate the sampling time and the sample number respectively.
Variables written with the letters V and q are used to indicate volume and flow respectively. The superscripts $in$, $out$, $u$, and $w$ indicate the inflow, outflow, control flow, and weir overflow respectively.
 \begin{figure}[h]
   \centering
     \includegraphics[width = 0.5\textwidth,trim={0.8cm 0.1cm 1.20cm 0.1cm},clip ]{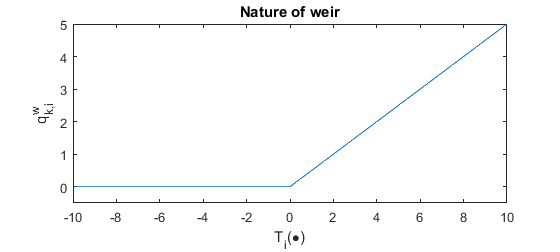}
     \caption{An illustration of the nature of weirs where the weir flow $q^w_{k,i}$ is zero when the switching function $T(\bullet)$ is negative, and following a given non-negative weir function $t_w(T)$, when the switching function is positive.}
      \label{fig:weir0}
\end{figure}

\section{Stochastic MPC with weir elements}
Systems with weirs or weir-like structures, such as sewer systems, have a binary nature originating from the weirs. The binary natures are shown in Fig. \ref{fig:weir0} for a linear weir function $t_w(T)$.  The binary nature can easily be observed, by noting that the flow is zero when the switching function $T_i(\bullet)$ is negative, and otherwise follows some given function depending on the switching function. The general deterministic formulation of MPC for systems with weirs can be formulated as below.
\begin{align}
J &= \min\limits_{\textbf{u}} f(\textbf{x},\textbf{u},\textbf{z}^{ref},\textbf{w},\textbf{q}^w)\label{eq:MPC}\\
\textbf{x}_{k+1} &=  h_{proc}(\textbf{x}_k,\textbf{u}_k, \textbf{w}_k,\textbf{q}^w_k), \quad \textbf{x}_0 = \textbf{x}_{ini}\label{eq:mpcP}\\
q^{w}_{k,i} &= \begin{cases}
t_{w,i}(T_i(\bullet)), \quad T_i(\bullet) \geq 0\\ 0
\end{cases}\quad \forall i \in \{1:N_w\}\label{eq:smpcW}\\
&\textbf{h}(\textbf{x}_k,\textbf{w}_k,\textbf{u}_k,\textbf{q}^w_k) = 0 \label{eq:MPC1}\\
&\textbf{g}(\textbf{x}_k,\textbf{w}_k,\textbf{u}_k,\textbf{q}^w_k) \leq \bar{\textbf{g}}\label{eq:MPC2}
\end{align}
Where $\textbf{x}$ corresponds to the system states, the control of the system is $\textbf{u}$, $\textbf{w}$ is the rain inflow into the system, and the weir flow $q^w_{k,i}$ corresponds to the ith weir element out of $N_w$ at time k, and is always non-negative, while $\textbf{x}_{ini}$ is the system's initial condition. 

As mentioned earlier, we will in this work consider the stochastic MPC method CC-MPC. Using the CC-MPC method to handle uncertainty, the cost function in (\ref{eq:MPC}) and the equality constraints in (\ref{eq:MPC1}) are rewritten as the expectation of the given function. The inequality constraints in (\ref{eq:MPC2}) are reformulated as the probability of the constraints holds true with a given probability. The process equation (\ref{eq:mpcP}) and weir definition (\ref{eq:smpcW}) can be substituted into the cost function and constraints so that the only state the system is explicitly depending on is the initial state $\textbf{x}_0$. Due to the presence of weirs with (\ref{eq:smpcW}), the resulting probability functions become meaningless as will be shown later. Therefore, we will reformulate the CC-MPC formulation, such that the inclusion of weir structures gives a sensible expression of the probabilistic formulation.

\subsection{Revised CC-MPC Formulation}
In our revised formulation of CC-MPC, we will formulate how to include weir structures in the probabilistic formulation, but we will also consider the feasibility of the program, as well as overflow determination. In the formulation of the cost function and the equality constraints, the approach utilized in standard CC-MPC can be reused as given below.
\begin{align}
J = \min\limits_{\textbf{u}}E\{ f(\textbf{x}_0,\textbf{u},\textbf{z}^{ref},\textbf{w},\textbf{q}^w)\}\\
0 =  E\{h(\textbf{x}_0,\textbf{u}, \textbf{w}, \textbf{q}^w)\}
\end{align}
Where $\textbf{q}^w$ is written for clarity of the presence of weirs.
The formulation of the probability constraints in CC-MPC can cover sets of constraints or individual constraints. We consider the latter in this work, where the approach to the reformulation of the inequality constraints depends on the specific constraint. If the constraint does not contain a weir element, meaning that no weir overflow is defined by this particular constraint, then the direct probabilistic approach from CC-MPC can be utilized to handle the uncertainty. Below is shown the probabilistic rewriting of the ith inequality constraint (\ref{eq:g}), into the quantile function-based constraint (\ref{eq:g1}), with arrows indicating the order of steps in the process.
\begin{align}
	g_i(\bullet) &\leq \bar g_i \label{eq:g}\\
	\rightarrow \quad Pr\{g_{i}(\bullet) &\leq \bar g_i\} \geq \alpha\\
	\rightarrow \quad \Phi^{-1}_{g_i(\bullet)}(\alpha) &\leq \bar g_i\label{eq:g1}
\end{align}
The above quantile function is based on the distribution of the constraint. Given the optimization variables are contained in the quantile, this is difficult to solve optimization-wise. Utilizing standardization of the constraint distribution, this can be simplified as shown in (\ref{eq:g2}), where the distribution is assumed defined purely by its expectation and variance. Such distributions include the normal distribution; therefore, we will utilize this assumption in the rest of this discussion of the reformulation of the CC-MPC. 
\begin{equation}\label{eq:g2}
 E\{g_i(\bullet)\} \leq \bar g_i - \sigma_{g_i(\bullet)}\Phi^{-1}(\alpha)
\end{equation}

If the constraint does define a weir overflow, then the direct probabilistic approach results in a meaningless probability. This is due to the weir element making the constraint intrinsically feasible, by counteracting the breaching of the constraint, as demonstrated below:
\begin{align}
g_i(\bullet,q^w_{k,i}) &\leq \bar g_i, \quad q^w_{k,i} = 0\\
g_i(\bullet,q^w_{k,i}) &= \bar g_i, \quad q^w_{k,i} > 0\\
\rightarrow Pr\{g_i(\bullet,q^w_{k,i}) &\leq \bar g_i \} = 1
\end{align}
where regardless of which parameters $\bullet$ the constraint depends on, the weir overflow will be depending on the same parameters so that the constraint holds.

For this reason, including the above constraint in the optimization formulation is redundant. In order for achieving a statistical bound on the overflow generation, we instead turn to the probability of keeping the weir overflow $q^w_{k,i}$ non-positive, where we can see this is related to its switching function $T(\bullet)$, as shown below.
\begin{align}
Pr\{q^w_{k,i} \leq 0\} &= Pr\{ T_i(\bullet) \leq 0 \} \geq \gamma \\
\rightarrow \quad E\{T_i(\bullet)\} &\leq -\sigma_{T_i(\bullet)}\Phi^{-1}(\gamma)\label{eq:Teq}
\end{align}
This allows us to formulate probabilistic constraints for both constraints with or without weir elements, as were shown in (\ref{eq:Teq}) and (\ref{eq:g2}).

\subsection{Feasibility}
In the above, we only considered handling the uncertainty such that a given solution would be feasible in the real system with known probability. This leads to probabilistic restrictions on the inequality constraints, but these restrictions will also lead to more rain scenarios causing infeasibility during computations. By utilizing slack variables with a suitable cost term in the cost function, we can restore the original feasibility of the constraints by the following approach, while keeping the probabilistic restrictions, when possible. 
\begin{align}
 E\{g_i(\bullet)\} &\leq \bar g_i + s_k  - \sigma_{g_i(\bullet)}\Phi^{-1}(\alpha)\label{eq:feas1}\\
E\{T_i(\bullet)\} &\leq c_k -\sigma_{T_i(\bullet)}\Phi^{-1}(\gamma)\label{eq:feas2}\\
0 &\leq s_k, c_k
\end{align}
Where the constraints without weirs are given by (\ref{eq:feas1}) and the weir defining constraints is given by (\ref{eq:feas2}). For the constraint without weirs, an extra constraint is necessary to represent the original constraint of the system:
\begin{equation}\label{eq:slack0}
	s_k  - \sigma_{g_i(\bullet)}\Phi^{-1}(\alpha) \leq 0
\end{equation}
Using the above versions of the constraints, the formulation of the optimization program for feasible CC-MPC can be written as the following:
\begin{align}
J = \min\limits_{\textbf{u},\textbf{c},\textbf{s}}E\{ &f(\textbf{x}_0,\textbf{u},\textbf{z}^{ref},\textbf{w},\textbf{q})\} + l(\textbf{c},\textbf{s})\\
0 &=  E\{h(\textbf{x}_0,\textbf{u}_k, \textbf{w}_k, \textbf{q}_k^w)\}\\
 E\{g_i(\bullet)\} &\leq \bar g_i + s_k  - \sigma_{g_i(\bullet)}\Phi^{-1}(\alpha)\\
E\{T_i(\bullet) \}&\leq c_k -\sigma_{T_i(\bullet)}\Phi^{-1}(\gamma)\\
s_k  &\leq \sigma_{g_i(\bullet)}\Phi^{-1}(\alpha)\\
0 &\leq s_k, c_k
\end{align}
where the additional function $l(c,s)$ in the cost functions is the cost term of the slack variables, penalizing their usage.

\subsection{ Overflow Approximation}
So far, we have been considering an optimization formulation with a dynamic description of the weir overflows included in it.
Given the nature of weir overflows being binary as seen in (\ref{eq:smpcW}) and therefore not being convex, the inclusion of the dynamic can lead the optimization program to be computational heavy. One approach to deal with this is to treat the weir overflows as additional optimization variables and penalize their utilization\cite{HFC}. Given that overflows cannot be negative, a constraint for this needs to be added. Another aspect is the determination of the value of the overflow for approximation; Given that we are minimizing the overflow, we need a constraint telling us the minimum size of the overflow. A fitting constraint for this is the original constraint containing the overflow, due to it being its very definition. We can utilize the expectation of this constraint, to achieve a description of the overflow size and still taken care of the uncertainty. Based on the added constraint shown below, our approximated overflow can be considered the expected overflow of the system in some sense. 
\begin{align}
E\{g_i(\bullet,\textbf{q}^w)\} &\leq \bar g_i\\
0\leq q^w_k
\end{align}
With the approximation approach we have utilized, we can formulate the optimization program as below. The cost function now includes a penalty term on the overflow variables. This term is a penalty on the accumulated overflow volumes at each sample in the predictions.
\begin{align}
J = \min\limits_{\textbf{u},\textbf{c},\textbf{s},\textbf{q}^w}E\{ f(\textbf{x}_0,\textbf{u}&,\textbf{z}^{ref},\textbf{w},\textbf{q})\} + l(\textbf{c},\textbf{s}) + \sum\limits_{k=0}^N \textbf{M}_{k}^T\textbf{q}^w_k\\
0 &=  E\{h(\textbf{x}_0,\textbf{u}_k, \textbf{w}_k, \textbf{q}_k^w)\}\\
 E\{g_i(\bullet)\} &\leq \bar g_i + s_k  - \sigma_{g_i(\bullet)}\Phi^{-1}(\alpha)\\
E\{T_i(\bullet) \}&\leq c_k -\sigma_{T_i(\bullet)}\Phi^{-1}(\gamma)\\
s_k  &\leq \sigma_{g_i(\bullet)}\Phi^{-1}(\alpha)\\
E\{g_i(\bullet,q^w_k)\} &\leq \bar g_i\\
0 &\leq s_k, c_k, q^w_k
\end{align}

\section{Model \& Cost}\label{sec:3}
\begin{figure}
   \centering
     \includegraphics[width = \textwidth,trim={2.40cm 24.0cm 0.40cm 0.0cm},clip ]{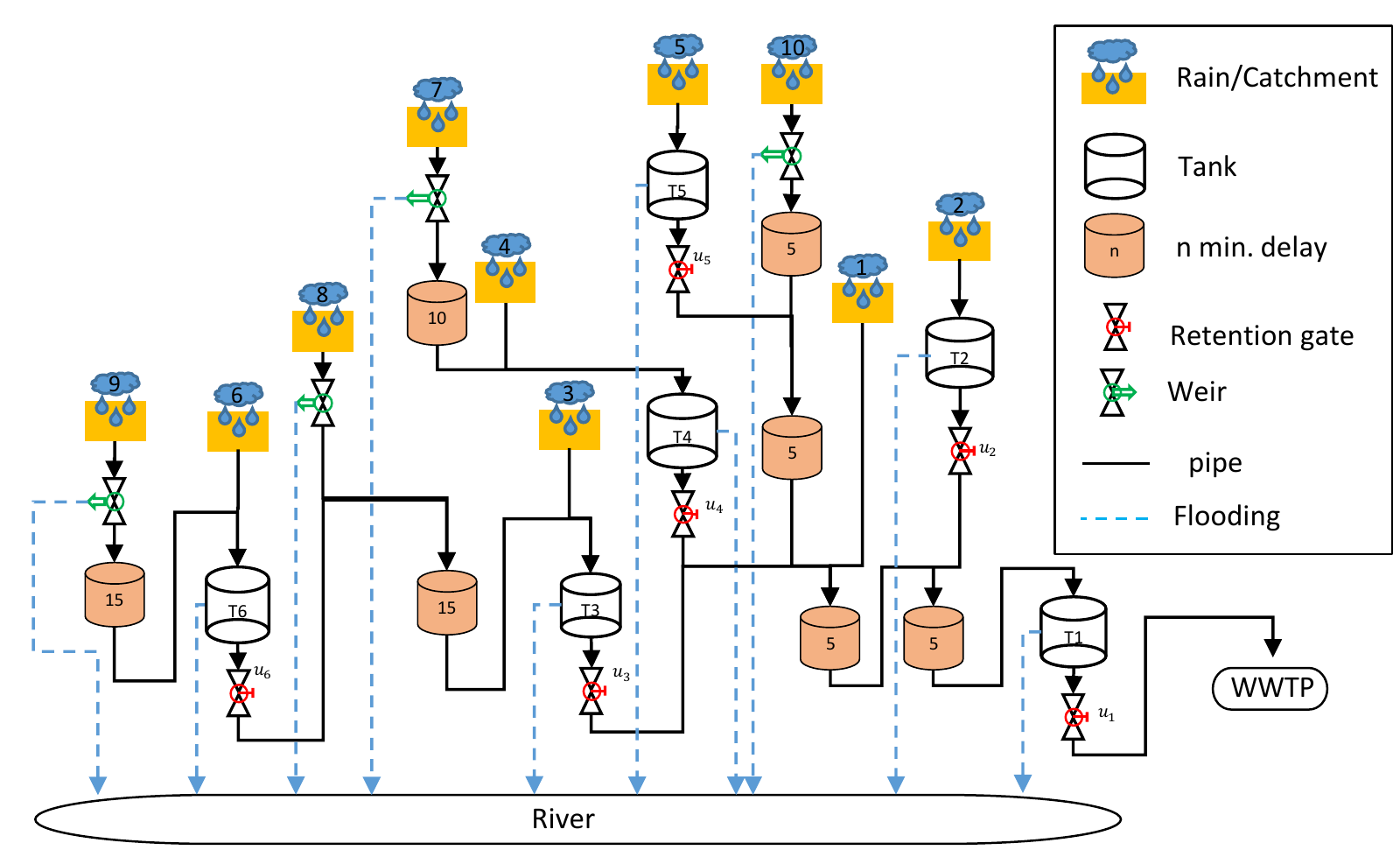}
     \caption{A schematic of the system model based on a linear version of the Astlingen Benchmark Network\cite{MMMU}. It shows the interconnections between tanks, pipes, and the environment. The parts tagged with a $u_i$ is controllable.}
      \label{fig:system}
\end{figure}

In this section, we will outline an example of the application of the revised CC-MPC formulation. For clarity, we will first outline the design model of the deterministic MPC followed by the stochastic counterpart. The system considered is a linear model of the Astlingen sewer network\cite{MMMU} illustrated in Fig. \ref{fig:system}. The Astlingen system consists of 10 catchment areas connected to a system of 6 controllable tanks and 4 independent weirs, all capable of flooding the nearby river by overflows.
For the cost function of the MPC given below, we will utilize a mix of linear and quadratic cost terms, including the overflow approximation approach, discussed previously\cite{HFC}\cite{JLS1}.
\begin{align} \label{eq:cost}
	J = \min\limits_{\textbf{u},\textbf{q}^w} \sum\limits_{k=0}^N &||\Delta \textbf{u}_k||_R^2 + \textbf{Q}^T\textbf{z}_k + \textbf{W}^T\textbf{V}^w_k\\
	\textbf{V}^w_k &= \sum\limits_{i=0}^k \textbf{q}^w_i
\end{align}
where the cost is minimized over an N step prediction horizon on the system, with a quadratic penalty of the control change $\Delta \textbf{u}$ and a linear cost on the output objective $\textbf{z}$. The output objectives correspond to the following objectives:
\begin{itemize}
\item maximizing flow to WWTP
\item minimizing flow to the environment
\end{itemize}
The third term $\textbf{W}^T\textbf{V}^w_k$ is a linear penalty on the accumulated overflow volume at time k.

The system can be considered to consist of tanks, pipes with weirs, and delay pipe elements. If the sizes of the delays are in multiples of the sampling time, then they can be considered a cascade of delays of the size of the sampling time, $D_{k,i}$. The dynamics of the tanks and delays are described by the following equations:
\begin{align}
	V_{k+1,i} &= V_{k,i} + \Delta T (q^{in}_{k,i}-q^{out,V}_{k,i}-q^w_{k,i})\\
	D_{k+1,i} &= q^{in}_{k,i}
\end{align}
The outflows of each element are described by the equations below, where V, P, and D indicate the type of element; tank volume, pipe flow, and delay flow respectively.
\begin{align}
	q^{out,V}_{k,i}&=  u_{k,i} \\
           q^{out,P}_{k,i}&= q^{in}_{k,i} - q^w_{k,i}\\
	q^{out,D}_{k,i}&= D_{k,i}
\end{align}
The inflow $q^{in}_{k,i}$ of the ith element, given below, are dependent on the connections of the elements in the system as shown in Table \ref{tab:inflow}. 
Where the ith tank is denoted by Ti, pipe i from catchment i by pi and the n minute delay to tank i is given by Ti:n.
\begin{equation}
q^{in}_{k,i} = w_{k,i}+\sum\limits_{j\in\mathcal{U}_i} u_{k,j}+\sum\limits_{j\in\mathcal{Q}^V_i} q^{out,V}_{k,j}+\sum\limits_{j\in\mathcal{Q}^P_i} q^{out,P}_{k,j} +\sum\limits_{j\in\mathcal{Q}^D_i} q^{out,D}_{k,j}\\
\end{equation}
Where the j denotes the flows of the subsets $\mathcal{U}_i$, $\mathcal{Q}^V_i$, $\mathcal{Q}^P_i$, $\mathcal{Q}^D_i$ of all control flows, tank outflows, pipe outflows, and delay outflows respectively. The variable $w_{k,i}$ indicates the rain inflow to the system part.
\begin{table}
\large
\caption{Inflows to the different elements of the systems}
\centering
\begin{tabular}{ll}
\begin{tabular}{|l|l|}
\hline
Element & Inflow \\
\hline
T1 & $q^{out,D}_{k,T1:5}$ \\
T2 & $w_{k,2}$ \\
T3 & $w_{k,3}+q^{out,D}_{k,T3:5}$\\
T4 & $w_{k,4}+q^{out,D}_{k,T4:5}$ \\
T5 & $w_{k,5}$ \\
T6 & $w_{k,6} + q^{out,D}_{k,T6:5}$ \\
  \hline
T3:5 & $q^{out,D}_{k,T3:10}$\\
T3:10 & $q^{out,D}_{k,T3:15}$\\
T3:15 & $q^u_{k,6} + q^{out,P}_{k,p8}$\\
\hline
T4:5 & $q^{out,D}_{k,T4:10}$\\
T4:10 &$q^{out,P}_{k,p7}$\\
\hline
\end{tabular}
&
\begin{tabular}{|l|l|}
\hline
Element & Inflow \\
\hline
p7 & $w_{k,7}$\\ 
p8 & $w_{k,8}$\\
p9 & $w_{k,9}$\\
p10 & $w_{k,10}$\\
\hline
T1:5 &$ q^u_{k,2}+ q^{out,D}_{k,T1:10}$ \\
T1:10 & $w_{k,1} + q^u_{k,3}+ q^u_{k,4} + q^{out,D}_{k,T1:15}$\\
T1:15 & $q^u_{k,5} + q^{out,D}_{k,T1:20}$\\
  T1:20 & $q^{out,P}_{k,p10}$\\
\hline
T6:5 & $q^{out,D}_{k,T6:10}$\\
T6:10 & $q^{out,D}_{k,T6:15}$\\
 T6:15 & $q^{out,P}_{k,p9}$\\
\hline
\end{tabular}
\end{tabular}
\label{tab:inflow}
\end{table}

The inequality constraints are formulated below, where the upper and lower limits of the tank volumes, the pipe outflow, control flow, and the weir overflows are stated for time k. The constraints are based on the individual elements i of the system. Not all of the constraints are applicable for all types of elements, e.g. (\ref{eq:Vcon}) are only applicable for tanks.
\begin{align}
	0\leq V_{k,i} -\Delta T q^w_{k,i}  &\leq \bar V_i\label{eq:Vcon}\\
	0 \leq q^{out,P}_{k,i}&\leq \bar q^{out,P}_{i}\\
	0 \leq q^{out,V}_{k,i} &\leq \beta_i V_{k,i} \label{eq:Vcon1}\\
	0 \leq u_{k,i} &\leq \bar u_{i}\label{eq:Vcon2}\\
	0 \leq q^w_{k,i}\label{eq:Vcon3}
\end{align}
Where $\beta$ is the volume-flow coefficient\cite{Singh1988}. 
The upper constraints of the tank volumes and pipe outflows are the definitions of the occurrence of overflow. With the corresponding switching function given by (\ref{eq:T}).
\begin{equation} \label{eq:T}
	T_i(\bullet) = \begin{cases}V_{k,i}-\bar V_i&\quad\text{(Tank)}\\ q^{in}_{k,i}-\bar q^{out,P}_i &\quad\text{(Pipe)}\end{cases}
\end{equation}

\subsection{Stochastic Model}\label{sec:stoch}
The revised formulation of CC-MPC with overflow handling presented earlier can now be applied to the system described above. The cost function of the revised CC-MPC can then be written as:
\begin{equation}
	J = \min\limits_{\textbf{u},\textbf{c},\textbf{s},\textbf{q}^w} \sum\limits_{k=0}^N E\{||\Delta \textbf{u}_k||_R^2 + \textbf{Q}^T\textbf{z}_k + \textbf{W}^T\textbf{V}^w_k\}  + \textbf{W}^T_c\textbf{c} + \textbf{W}^T_s\textbf{s}
\end{equation}
Both the rewritten cost function and the later inequality constraints depend on the expectation of the system's subpart equations. Where the expected tank volume, delay flow, element inflow, tank outflow, pipe outflow, and delay outflow are given by (\ref{eq:EV})-(\ref{eq:EqoutD}) respectively.
\begin{align}
	\label{eq:EV}E\{V_{k+1,i}\} &= E\{V_{k,i}\} + \Delta T (E\{q^{in}_{k,i}\}-E\{q^{out,V}_{k,i}\}-q^w_{k,i}\\
	\label{eq:ED}E\{D_{k+1,i}\} &= E\{q^{in}_{k,i}\}\\
	E\{q^{in}_{k,i}\} &= E\{w_{k,i}\}+\sum\limits_{j\in\mathcal{U}_i} u_{k,j}+\sum\limits_{j\in\mathcal{Q}^V_i} E\{q^{out,V}_{k,j}\} + \sum\limits_{j\in\mathcal{Q}^P_i} E\{q^{out,P}_{k,j}\} +\sum\limits_{j\in\mathcal{Q}^D_i}E\{ q^{out,D}_{k,j}\} \label{eq:Eqin}\\
	\label{eq:Eqout}	E\{q^{out,V}_{k,i}\} &=  u_{k,i}\\
	E\{q^{out,P}_{k,i}\} &= E\{q^{in}_{k,i}\} - q^w_{k,i}\label{eq:EqoutP}\\
	\label{eq:EqoutD}E\{q^{out,D}_{k,i}\} &= E\{D_{k,i}\}
\end{align}
In the following paragraphs, the formulation of each inequality constraint is given in the context of the corresponding subpart. The resulting formulation of the lower constraint of the tank volume is given by:
\begin{align}\label{eq:LVC}
	\sigma_{V_{k,i}}\Phi^{-1}(\alpha_j) - s_{j,k} & \leq E\{V_{k,i}\} - \Delta T q^w_{k,i}\\
	0 \leq s_{j,k} &\leq\sigma_{V_{k,i}}\Phi^{-1}(\alpha_j)
\end{align}
where j indicates the specific constraint. The probabilistic formulation of the upper constraint of the tank volume is then defined by the switching function (\ref{eq:T}) as written in (\ref{eq:UVC}). The constraint for overflow approximation is given by (\ref{eq:UVCA}).
\begin{align}\label{eq:UVC}
	E\{V_{k,i}\} &\leq \bar V_i -\sigma_{V_{k,i}}\Phi^{-1}(\gamma)+ c_k\\
	E\{V_{k,i}\} -\Delta T q^w_{k,i}  &\leq \bar V_i\label{eq:UVCA}\\
		0&\leq c_k
\end{align}
From (\ref{eq:Eqout}), we know that the tank outflows are controlled and therefore deterministic, this gives us that the lower limit is the same as in (\ref{eq:Vcon1}).
For the upper limit, the probabilistic formulation is given by (\ref{eq:CUO}) and (\ref{eq:CUO1}).
\begin{align}\label{eq:CUO}
	u_{k,i} &\leq \beta E\{V_{k,i}\}-\beta \sigma_{V_i}\phi^{-1}(\alpha_j) + s_{j,k}\\
	0 &\leq s_{j,k}\leq \beta \sigma_{V_i}\phi^{-1}(\alpha) \label{eq:CUO1}
\end{align}
The limits on the pipes with weirs are given by (\ref{eq:Pcon}) and (\ref{eq:Pcon1}) for the lower limit, and by (\ref{eq:Pcon2})-(\ref{eq:Pcon3}) for the upper limit.
\begin{align}
	\label{eq:Pcon}\sigma_{q^{out,P}_{k,i}}\Phi^{-1}(\alpha_j)- s_{j,k} &\leq E\{q^{out,P}_{k,i}\} \\
	\label{eq:Pcon1}0\leq s_{j,k}&\leq\sigma_{q^{out,P}_{k,i}}\Phi^{-1}(\alpha)
\end{align}
\begin{align}
	\label{eq:Pcon2}E\{q^{in}_{k,i}\} - q^w_{k,i}&\leq \bar q^{out,P}_{i}\\
	E\{q^{in}_{k,i}\}&\leq \bar q^{out,P}_i + c_{j,k} -\sigma_{q^{in}_{k,i}}\Phi^{-1}(\gamma_j)\\
	\label{eq:Pcon3}0&\leq c_{j,k}
\end{align}
Given that there is, per definition, no uncertainty in optimization variables, the constraints on the control and weir overflow are deterministic and are therefore the same as in (\ref{eq:Vcon2}) and (\ref{eq:Vcon3}).

\subsection{Benefits and costs}
The utilization of the approximation method discussed above has some significant drawbacks as previously discussed in \cite{JLS1}. The main drawbacks are the loss of design freedoms in the weighting of the cost function. These come from the extra weights on the aggregated overflow volume has to be relatively higher than the main terms of the cost functions, and have hierarchically weightings depending on their relative placement in the systems. 

These design restrictions on the weightings limit the flexibility of the control with regards to the planning of overflow countermeasures. While the revised CC-MPC does not change these drawbacks, it might give a possible remedy for the hierarchical weightings requirement. If a given weir overflow in the system is more attractive to society than weir overflow further down the system (e.g. downstream is a bathing area). Then by having a higher probability guarantee ($\alpha,\gamma$ in (\ref{eq:g2}), (\ref{eq:Teq}), and section \ref{sec:stoch}  ) on the downstream part than on the specific upstream overflow, the downstream constraints will be less likely to cause an overflow, if it is possible to avoid.
\\ \\
The revised CC-MPC formulation has the drawbacks of introducing more optimization variables and inequality constraints, even without the weir overflow approximation. These drawbacks arise from the conserved feasibility through the slack variables and the constraints on these for elements without weirs. The revised CC-MPC also has the clear benefits of conserving feasibility but more importantly giving statistical constraints on overflow generation, similar to the CC-MPC formulation for systems without internal overflow description.

\subsection{Variance of Constraints}
Given the assumption of the variance of the probabilistic constraints exist, and that the probabilistic constraints are scalar, the variance of each constraint is also scalar. We can utilize this feature to derive a computationally simple method for computing the variance for the constraints. Firstly, we need to define the variance of each constraint.
\begin{align}
	\label{eq:sigV} \sigma^2_{V_{k,i}} &=  \sigma^2_{V_{k-1,i}} +  \Delta T^2\sigma^2_{q^{in}_{k-1,i}}\\
	\label{eq:sigD}\sigma^2_{D_{k,i}}& =\sigma^2_{q^{in}_{k-1,i}}\\
	\label{eq:sigP} \sigma^2_{q^{out,P}_{k,i}} &= \sigma^2_{q^{in}_{k,i}}\\
	\label{eq:sigD2} \sigma^2_{q^{out,D}_{k,i}} &= \sigma^2_{D_{k,i}}
\end{align}
\begin{align}
	\label{eq:sigIn}\sigma^2_{q^{in}_{k,i}}& =\sigma^2_{w_{k,i}} +\sum\limits_{j\in\mathcal{Q}^P_i}\sigma^2_{q^{out,P}_{k,j}} +\sum\limits_{j\in\mathcal{Q}^D_i} \sigma^2_{q^{out,D}_{k,j} }	
\end{align}
where each source of uncertainty is assumed to be independent both temporally and spatially. This gives equations for the variances, which is a linear model of the initial state variance and the rain inflow variances as shown in (\ref{eq:Var}). Utilizing this, all of the constraints discussed above can be combined into a matrix inequality covering the entire prediction horizon as shown in 
(\ref{eq:Var1})-(\ref{eq:Var2}).
\begin{align}
	\label{eq:Var}\pmb{\sigma}^2 &= \Theta\sigma^2_{\textbf{V}_0} + \Gamma\sigma^2_{\textbf{w}}\\
\label{eq:Var1}	\Omega_uU+\Omega_{q^w}q^w&\leq \Omega_{const} + \Omega_sS + \Omega_cC + \Omega_I\sigma_{Diag}\Phi^{-1}(\gamma)\\
\Omega_{const} &= \Omega + \Omega_xE\{\textbf{x}_0\} + \Omega_wE\{\textbf{W}\}\\
\sigma_{Diag} & = \sqrt{diag(\pmb{\sigma}^2)}\in \mathcal{R}^{nxn}\label{eq:Var2}
\end{align}

\section{Results \& Discussion}
In the previous sections, we have introduced the revised CC-MPC formulation, in this section; we will focus on analyzing the difference between the performance of the revised CC-MPC and the classical deterministic MPC applied to the Astlingen model introduced earlier. In the simulations, the examples of the MPC designs given in section \ref{sec:3} are used with a prediction horizon of a 100 min., where the weights of each objective in the cost function have the following values; $2$ for minimizing flow to nature, $-1$ for maximizing flow to WTTP, and $0.01$ for the change in control flow.
The higher the absolute weight is the higher the priority, while a negative weight indicates maximization, instead of minimization in the objective.  The weighting of the accumulated overflow volume is given in table \ref{tab:2}, where it can be seen the weights vary accordingly to the placements of the overflows in the system, as described in \cite{JLS1}. The usages of the slack variables are weighted uniformly with 100.

\begin{figure}
  \centering
  \begin{minipage}[b]{0.48\textwidth}
    \includegraphics[width = \textwidth,trim={0.8cm 0.1cm 1.20cm 0.1cm},clip ]{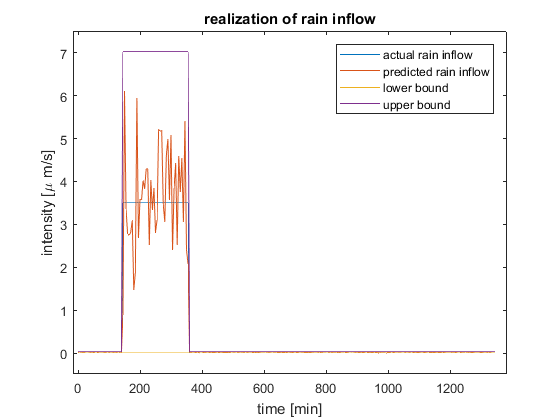}
     \caption{Realization of the rain forecast of a rain scenario with 3.5 $\mu m/s$ intensity and a three and a half-hour rain duration. Showing the uncertain prediction around the actual step inflow.}
      \label{fig:weather}
  \end{minipage}
  \begin{minipage}[b]{0.48\textwidth}
    \includegraphics[width = \textwidth,trim={0.8cm 0.1cm 1.20cm 0.1cm},clip ]{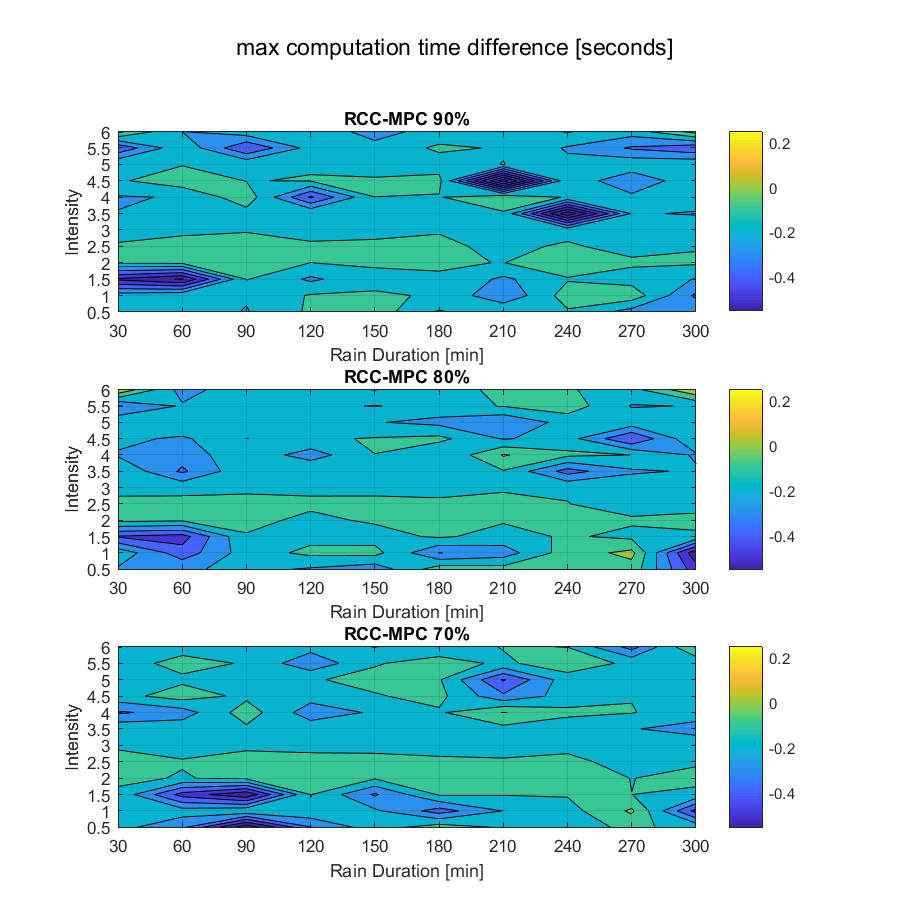}
\caption{The difference of Maximum computation time during each rain scenario simulation of the perfect MPC and the revised CC-MPC computed as MPC - revised CC-MPC}
      \label{fig:compu}
  \end{minipage}	
\end{figure}

Several scenarios with varying parameters have been run during the simulations. The profiles for the rain inflows in simulations were all step rains, where the rain intensity was varied from 0.5 to 6 $\mu m/s$, and the rain duration varying from a half-hour to five hours, with 0.5 $\mu m/s$ and half-hour intervals.  For the revised CC-MPC, the probability guaranty was equal across all constraints and was varied between scenarios, with values of $90\%$, $80\%$, and $70\%$ respectively. The deterministic MPC is assumed to have perfect forecasts of the rain inflow, while the revised CC-MPCs are operating with uncertainties following a truncated Gaussian distribution, with the expectation being the actual rain inflow. The size of the uncertainty for the CC-MPC (the standard variation $\sigma$), was chosen as a third of the expectation plus a constant deviation of $0.01 \mu m/s$, to avoid zero uncertainty. The truncated distribution of the uncertainty was assumed to be non-negative and below three $\sigma$ above the expectation, resulting in all realizations of the inflows to be within two times the expected non-zero value.
A realization of a rain scenario can be seen in Fig. \ref{fig:weather}, with the actual rain and bounds on the uncertainty, included.

\begin{table}
\caption{Cost function weighting of accumulated overflow volume \textbf{W}, showing a higher cost for upstream elements.}
\centering
\begin{tabular}{|l|l|l|l|l|l|l|l|l|l|}
\hline
T1 & T2 & T3 & T4 & T5 & T6 & p7 & p8 & p9 & p10\\
\hline
1000 & 5000 & 5000 & 5000 &5000 & 10000& 10000 & 10000 & 15000 & 5000\\
\hline
\end{tabular}
\label{tab:2}
\end{table}
\subsection{Computation Time}
From Fig. \ref{fig:compu}, we can observe the difference in the maximum computation time between the deterministic MPC and the revised CC-MPC with the three chosen probability guarantees.  It can be observed that in general, the revised CC-MPCs are around 0.1 seconds slower than the deterministic MPC, while occasional computations performer faster or slower for both types of controllers due to numerical variations.

\subsection{Weir Overflow}
\begin{figure}
  \centering
  \begin{minipage}[b]{0.48\textwidth}
    \includegraphics[width = \textwidth,trim={0.8cm 0.1cm 1.20cm 0.1cm},clip ]{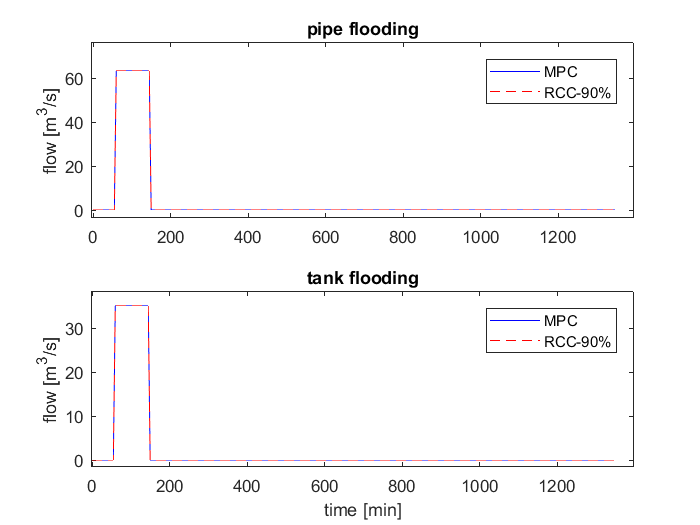}
     \caption{The total temporal overflow of a representative simulation}
      \label{fig:weir}
  \end{minipage}
  \begin{minipage}[b]{0.48\textwidth}
\includegraphics[width = \textwidth,trim={0.8cm 0.1cm 1.20cm 0.1cm},clip ]{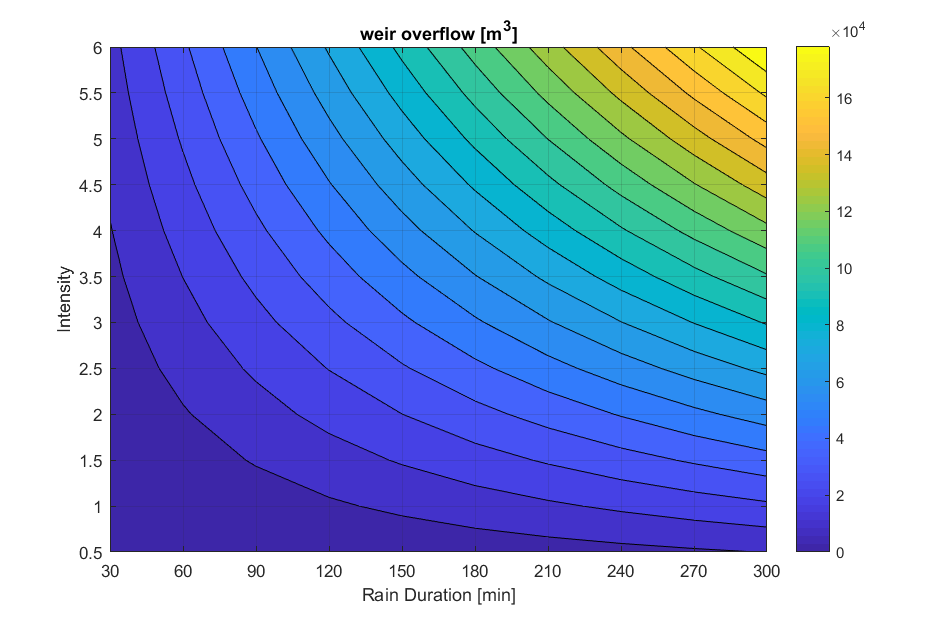}
     \caption{Total overflow volume of MPC with perfect knowledge}
      \label{fig:weir2}
  \end{minipage}	
\end{figure}

\begin{figure}[]
  \centering
  \begin{minipage}[b]{0.48\textwidth}
        \includegraphics[width = \textwidth,trim={0.8cm 0.1cm 1.20cm 0.1cm},clip ]{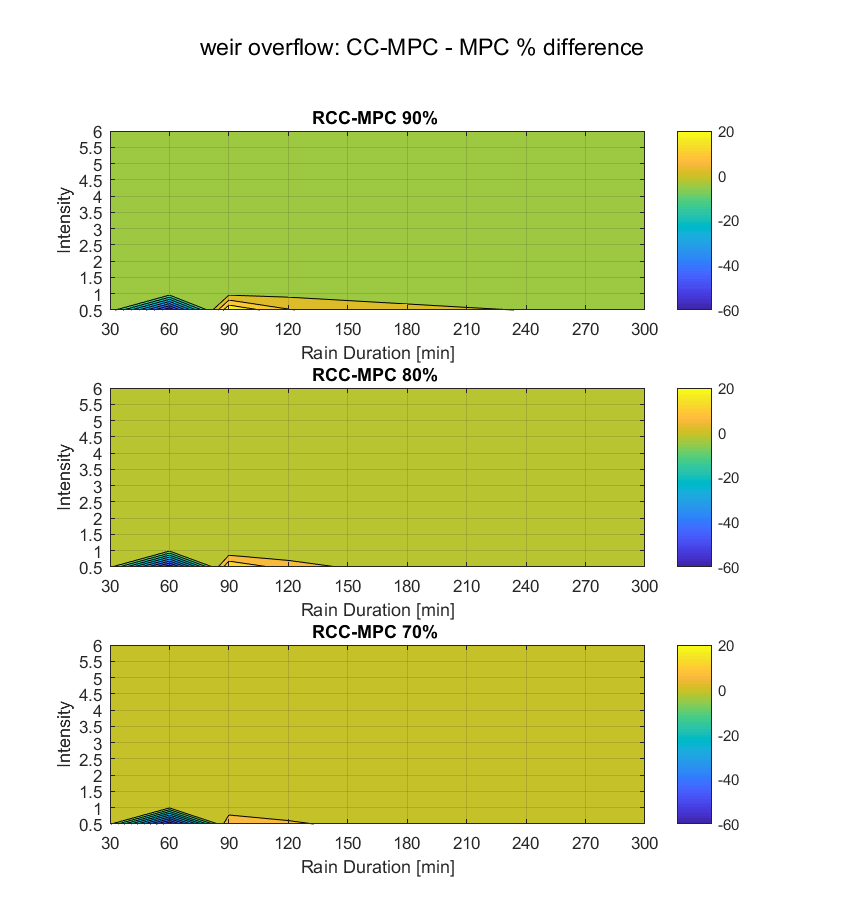}
     \caption{Percentage difference in total overflow experience, between revised CC-MPC and MPC with perfect knowledge}
      \label{fig:weir1}
  \end{minipage}
  \begin{minipage}[b]{0.48\textwidth}
    \includegraphics[width = \textwidth,trim={0.8cm 0.1cm 1.20cm 0.5cm},clip ]{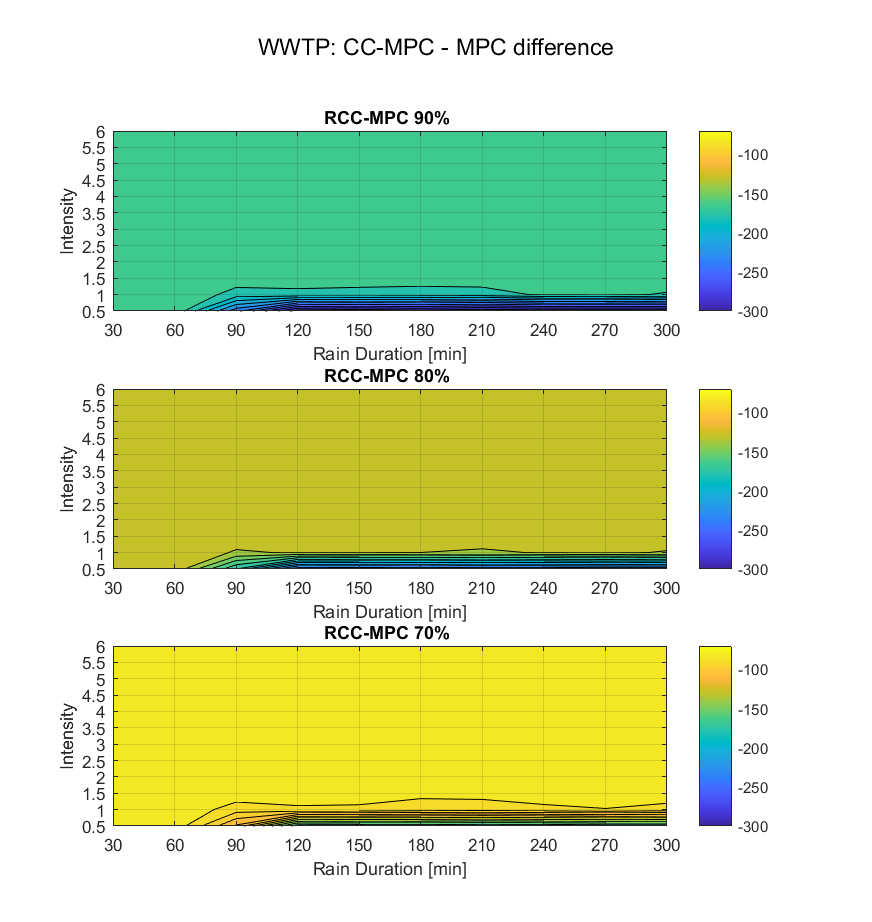}
     \caption{The difference in wastewater volume sent to the treatment plants, between the revised CC-MPC and the MPC with perfect knowledge}
      \label{fig:wttp1}
  \end{minipage}	
\end{figure}

In this part, we focus on the results of the simulation to do with weir overflows. In fig. \ref{fig:weir}, one of the simulations results is shown, showing the overflow over time for both the MPC with perfect forecast and the revised CC-MPC with $90 \%$ probability bound on the constraints. We can observe that the experienced overflow of the system is identical between the two controllers. This is further supported by the percental difference between the controllers for each rain scenario, shown in Fig. \ref{fig:weir1}. Here we can see that in general, the difference is approximately zero, but also that a few scenarios have larger differences. These differences are due to the conservatism of the CC-MPC when the end of the rain cannot be seen within the prediction horizon, giving the CC-MPC a better start position than the MPC for the next time step. The total volume of weir overflow of the deterministic MPC can be seen in Fig. \ref{fig:weir2}. By comparison to the percental differences before, we can see that the larger differences occurred, when the overflow volume is small for the MPC, making small divergences big in percentages.

\subsection{WWTP}
\begin{figure}
  \centering
  \begin{minipage}[b]{0.48\textwidth}
        \includegraphics[width = \textwidth,trim={0.8cm 0.1cm 1.20cm 0.1cm},clip ]{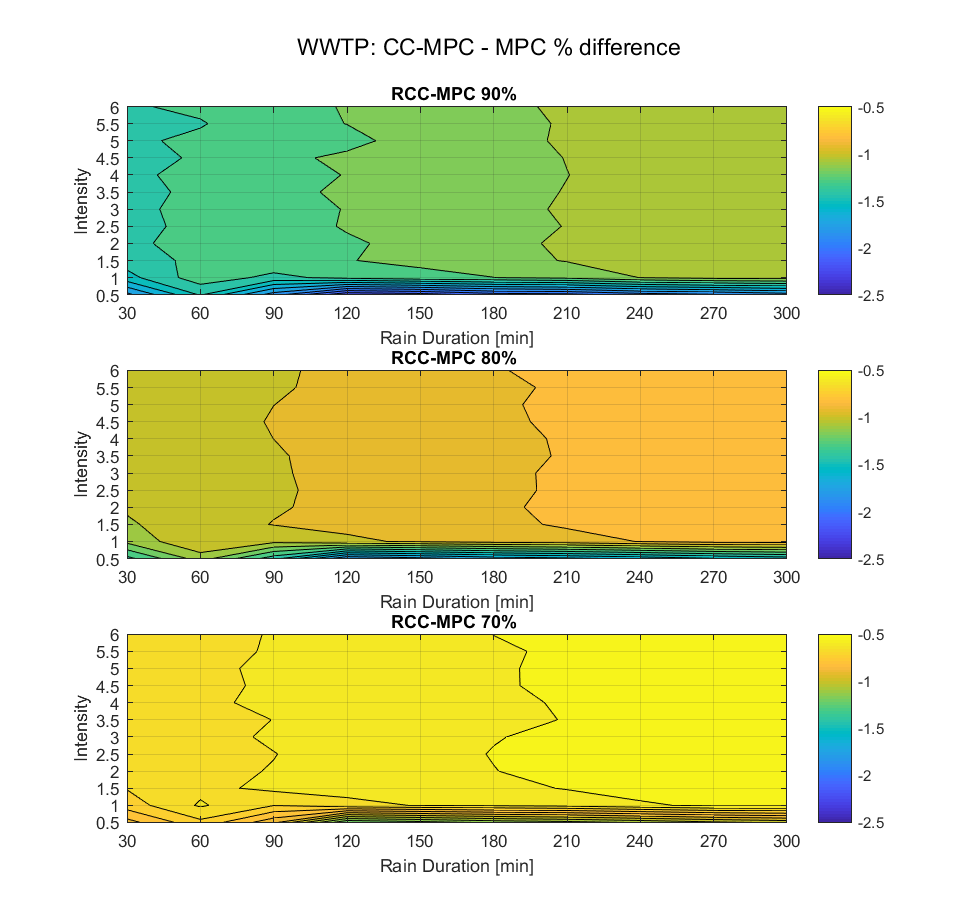}
     \caption{The percentage difference in wastewater volume sent to the treatment plants, between the revised CC-MPC and the MPC with perfect knowledge}
      \label{fig:wttp2}
  \end{minipage}
  \begin{minipage}[b]{0.48\textwidth}
    \includegraphics[width = \textwidth,trim={0.8cm 0.1cm 1.20cm 0.1cm},clip ]{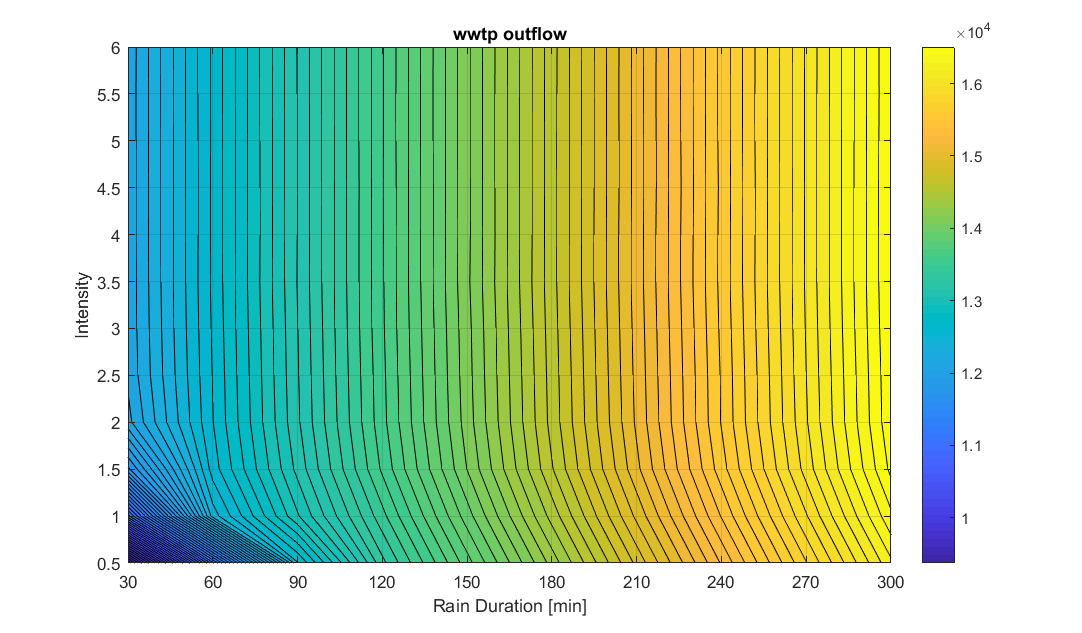}
     \caption{The total outflow volume of the deterministic MPC with perfect knowledge.}
      \label{fig:wwtp}
  \end{minipage}	
\end{figure}

In this part, we focus on the results of the simulation to do with the amount of water sent to the wastewater treatment plant of the system.
In Fig. \ref{fig:wttp1} and Fig. \ref{fig:wttp2}, we can observe the volumetric difference and the percentage difference in wastewater sent to the treatment plant respectively. 

We can see from the volume difference, that the deterministic MPC has an outflow, which is generally larger than the outflows of the revised CC-MPC, by somewhat constant volume bias around 120 $m^3$ depending on the probability bound.  We can observe from the percentage difference that while the bias is constant, the percentage difference primarily decreases with the duration of the rain, and not with the intensity of the rain. We can further see that the decrease in percentage difference corresponds to the increase  in the outflow, depicted for the deterministic MPC in Fig. \ref{fig:wwtp}

\subsection{Scenario example}
\begin{figure}
   \centering
     \includegraphics[width = 0.45\textwidth,trim={0.8cm 0.1cm 1.20cm 0.1cm},clip ]{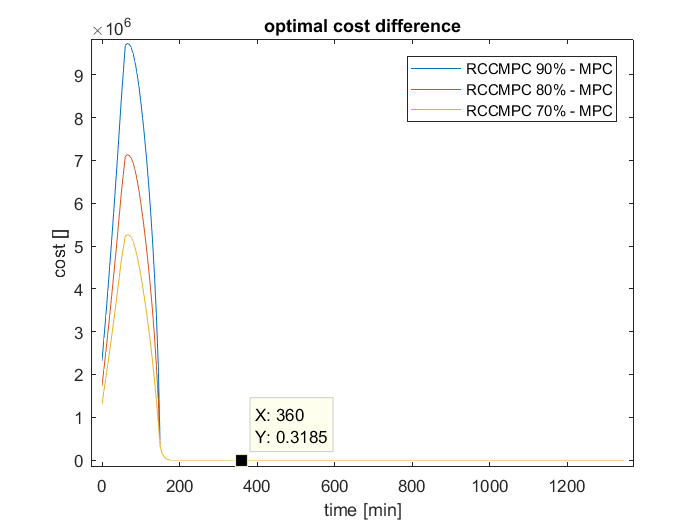}
     \caption{The difference in optimal cost function value between the revised CC-MPC and the MPC with perfect knowledge}
      \label{fig:realcost}
\end{figure}

In this section, we will focus on a representative simulation and the operational behavior across the system. The corresponding realization of the rain and the total overflow historic were already shown earlier in Fig. \ref{fig:weather} and Fig. \ref{fig:weir}. From Fig. \ref{fig:realcost}, we can observe the computational difference given as the cost difference. Here we can observe that the cost of the revised CC-MPC is always higher than for the deterministic MPC, with revised CC-MPC having a constant additional cost. It can also be observed that the revised CC-MPC with the highest probability bound has the highest cost, as one would expect.

\begin{figure}
  \centering
  \begin{minipage}[b]{0.48\textwidth}
        \includegraphics[width = \textwidth,trim={0.8cm 0.1cm 1.20cm 0.1cm},clip ]{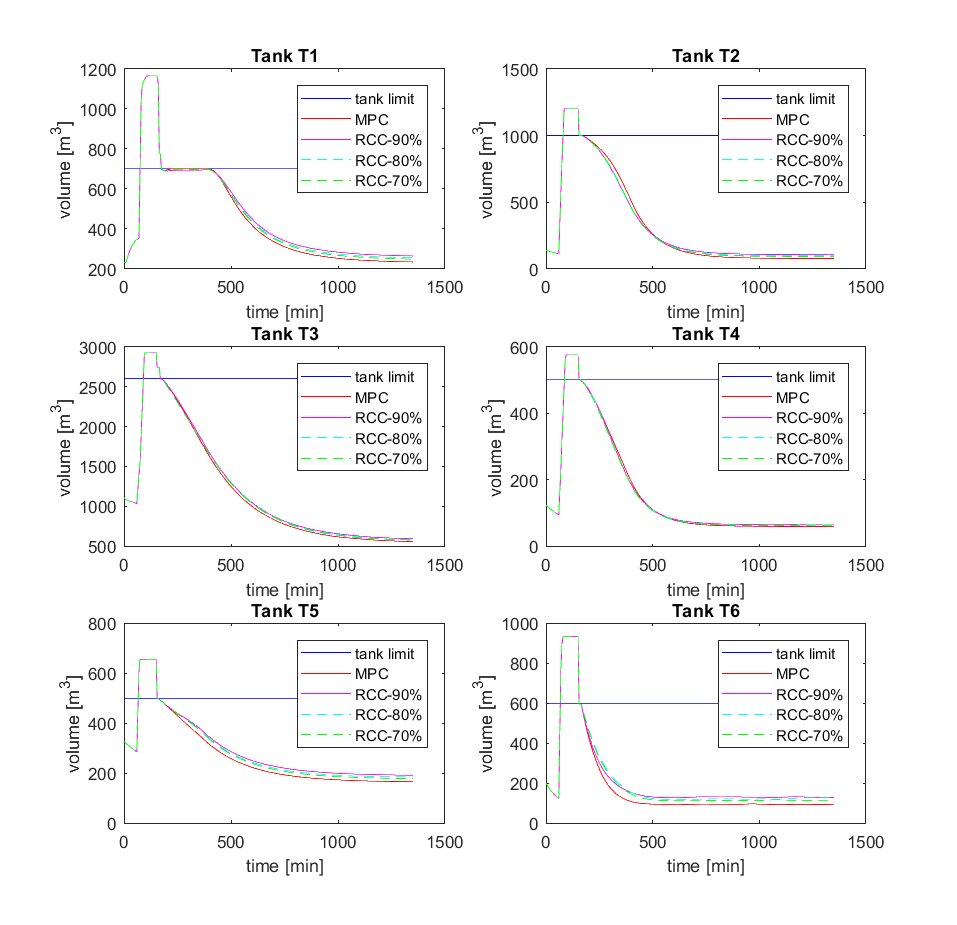}
     \caption{The volume historic of all six tanks for the deterministic MPC and the three revised CC-MPCs, with the addition of the physical tank limits. The blue lines indicate the tank limits, the red lines are the volume of the deterministic MPC, while the purple, green and cyan lines are the volume of the three revised CC-MPCs.}
      \label{fig:realtank}
  \end{minipage}
  \begin{minipage}[b]{0.48\textwidth}
    \includegraphics[width = \textwidth,trim={0.8cm 0.1cm 1.20cm 0.1cm},clip ]{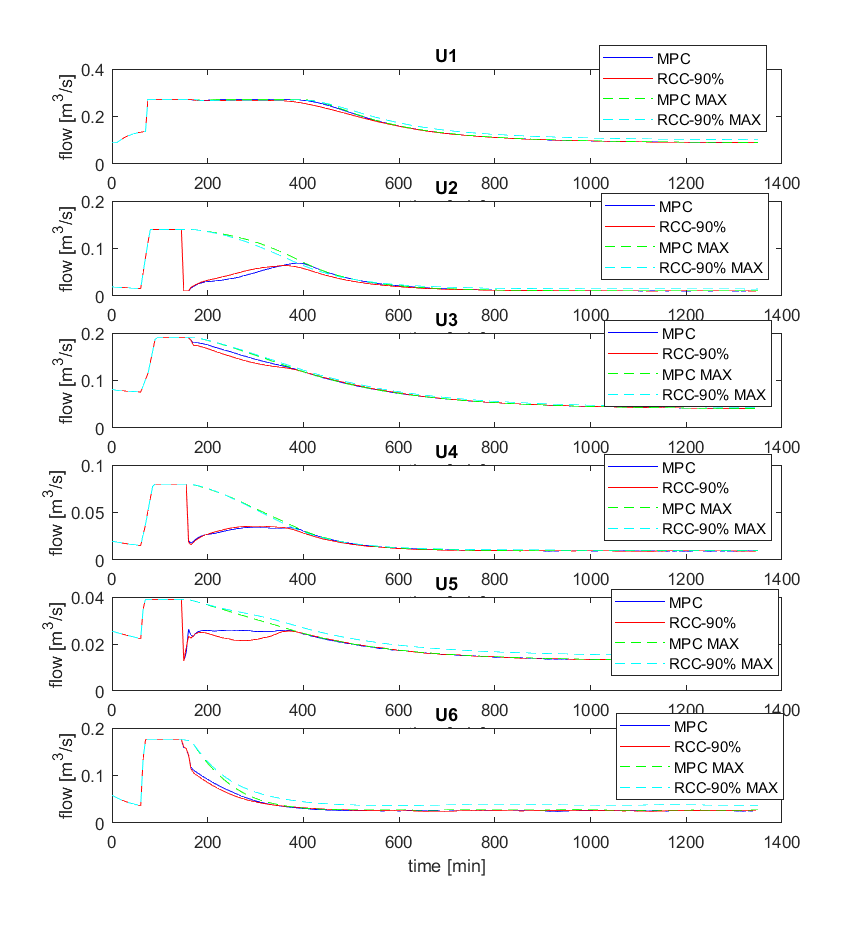}
     \caption{The control flow histories with the individual upper constraints marked for both the deterministic MPC and the revised CC-MPC with 90$\%$ probability bound. The blue line and the red line are the actual control flows of the MPC and revised CC-MPC respectively, while the green and cyan lines are the respective individual upper bounds. }
      \label{fig:realcon}
  \end{minipage}	
\end{figure}

The tank operational behavior of the system can be observed in Fig. \ref{fig:realtank}, with the tank outflow controllers displayed in Fig. \ref{fig:realcon}. From the tank volumes, we can once again see that the different MPCs agree on the optimal amount of volume exceeding the tanks.
We can also observe that the revised CC-MPCs find a steady-state volume, which is higher than the steady-state volume of the deterministic MPC.
This can further be observed by the control flows, where we can see that the control flows of both the deterministic MPC and the revised CC-MPC with $90 \%$ probability bound stay below the individual physical control constraint bounds of the control flows. We can see that the deterministic MPC, in general, operates slightly higher than the revised CC-MPC, and as expected can operate on the constraint bound. While the graph only depicts individual physical control constraint bounds, it is still interesting to note how far the steady-state operation of the revised CC-MPC is operating from the constraint bounds, due to the stochastic restraints. It can also be noted that the difference in operation between the two MPCs, first occurs after the rain has ended and not before, where rain was forecasted to happen. This indicates the difference is due to cost priority of the steady-state operation.

\section{CONCLUSIONS}
In this paper, we have presented a revised formulation of Chance-Constrained MPC (CC-MPC) inspired for application in sewer networks.
The main aspect of the reformulation focuses on preserving feasibility and introducing overflow handling of binary structures. 
The mathematical formulation and reasoning behind the revised CC-MPC has been stated and applied to a model of the Astlingen sewer system for testing the method.
A comparison of the performance of the revised CC-MPC and deterministic MPC was based  on simulations with idealized step rain inflows as the perturbation of the Astlingen system.
From the results of the simulations, it was shown that the weir overflow avoidance of the revised formulation provides similar results as the MPC with perfect forecast.
Indicating the revised CC-MPC as an alternative to MPC, when a perfect forecast is not achievable.
The results also showed a trade-off with regards to the worst-case computation time, which in general increased slightly. 

\bibliography{CCMPC_Sewer_formulation_preprint}

\end{document}